# Layer-dependence of dielectric response and water-enhanced ambient degradation of highly-anisotropic black As

Hwanhui Yun[1]*, Supriya Ghosh[1], Prafful Golani[2], Steven J. Koester[2], K. Andre Mkhoyan[1]*

[1]*Department of Chemical Engineering and Materials Science, University of Minnesota, Minneapolis, MN 55455*

[2]*Department of Electrical and Computer Engineering, University of Minnesota, Minneapolis, MN 55455*

**Corresponding Authors**

*E-mail: yunxx133@umn.edu (HY), mkhoyan@umn.edu (KAM)

**Abstract**

Black arsenic (BAs) is a van der Waals layered material with a puckered honeycomb structure and has received increased interest due to its anisotropic properties and promising performance in devices. Here, crystalline structure, thickness-dependent dielectric responses, and ambient stability of BAs nanosheets are investigated using STEM imaging and spectroscopy. Atomic-resolution HAADF-STEM images directly visualize the three-dimensional structure and evaluates the degree of anisotropy. STEM-EELS is used to measure the dielectric response of BAs as a function of the number of layers. Finally, BAs degradation under different ambient environments is studied highlighting high sensitivity to moisture in the air.

**Introduction**

Study of two-dimensional (2D) materials has been one of the active fields in materials science in last decade, leading advances in many research areas ranging from synthesis and characterizations to discovery of new physical phenomena and applications. Among various 2D materials, elemental black phosphorus (BP) occupies a unique position with attractive properties such as strong in-plane anisotropy, high carrier mobility, high-sensitivity of band structure to number of layers, and tunability of the bandgap spanning from near- to mid-infrared.[1-4] Another recently discovered and promising elemental 2D material from pnictogen is black arsenic (BAs) that has comparable characteristics to BP.[4-7] Earlier theoretical studies have predicted a highly anisotropic and tunable electronic structures for BAs.[8, 9] From the experimental side, recently Chen et al. showed the existence of extremely high in-plane anisotropy in electrical and thermal transport properties, while Kandemir et al. reported high mechanical and vibrational anisotropy in BAs.[5, 7] Thickness-dependent changes in the properties of BAs have also been explored. For example, Zhong et al. showed a layer number-dependent changes in the electronic band structures of BAs allowing adjustable carrier transport for BAs-based devices.[6]

Even at this early stage, BAs has already shown favorable properties and a potential to be one of the building block materials for advanced opto-electronic devices.[6] To better utilize this new 2D material and to maximize its performance, study of the fundamental physical and chemical properties, including its stability, is essential. For instance, poor stability of BP at ambient conditions gives a cautionary example for the importance of this topic. In case of BP, the ambient degradation includes formation of substance comprised of phosphorus oxides ($P_xO_y$) and condensed $H_2O$ and collapse of the layered atomic structure, which considerably restricts the practicality of using BP for opto-electronic devices.[10-19] Understanding and improving the ambient stability of BAs will determine its practical utility as well as limitations.

In this paper, a detailed analysis of structural and electronic properties of exfoliated BAs using analytical scanning transmission electron microscopy (STEM) is presented. Atomic-resolution high-angle annular dark-field (HAADF)-STEM images from different crystalline orientations were acquired and used to evaluate the degree of structural anisotropy of BAs. It is also shown that in case of few-layer-thick BAs, plan-view atomic-resolution HAADF-STEM images can be used to identify the exact number of layers in the nanosheet. In addition, electron energy-loss spectroscopy (EELS) was used to measure the features of the electronic structures of a BAs flake and changes in its dielectric response as a function of the number of layers. Finally, the stability of exfoliated BAs flakes has been examined under various ambient conditions, which allowed identification of the key degradation-enhancing parameters and suggested solutions for improving the long-term stability.

**Results and discussion**

*Atomic structure and anisotropy*

A plan-view HAADF-STEM image and energy dispersive X-ray (EDX) elemental maps of a BAs flake from a freshly prepared sample are shown in Figure 1. Low-magnification HAADF-STEM images from exfoliated flakes, as presented in Figure 1a, show presence of regions with different thicknesses – the brighter the HAADF signal, the thicker the region – with relatively sharp step edges between the regions. EDX elemental maps obtained from the flake confirm chemical composition of the flake as nearly-pure As without detectable impurities (impurity level in these flakes is less than 1 at.%, see supplementary information (SI), Figure S1). Cross-sectional

HAADF-STEM images of a BAs flake in low- and high-magnification were also obtained (Figure 1b) showing overall quality of the flakes and van der Waals layered crystalline lattice in the [101] direction. It should be noted that some flakes have a planar defect composed of a-few-atom-thick Pb, which is likely formed during synthesis of BAs crystal (see SI Figure S2).[38, 39] All other experiments and analysis, except for the one shown in SI Figure S2, were performed using defect-free BAs flakes.

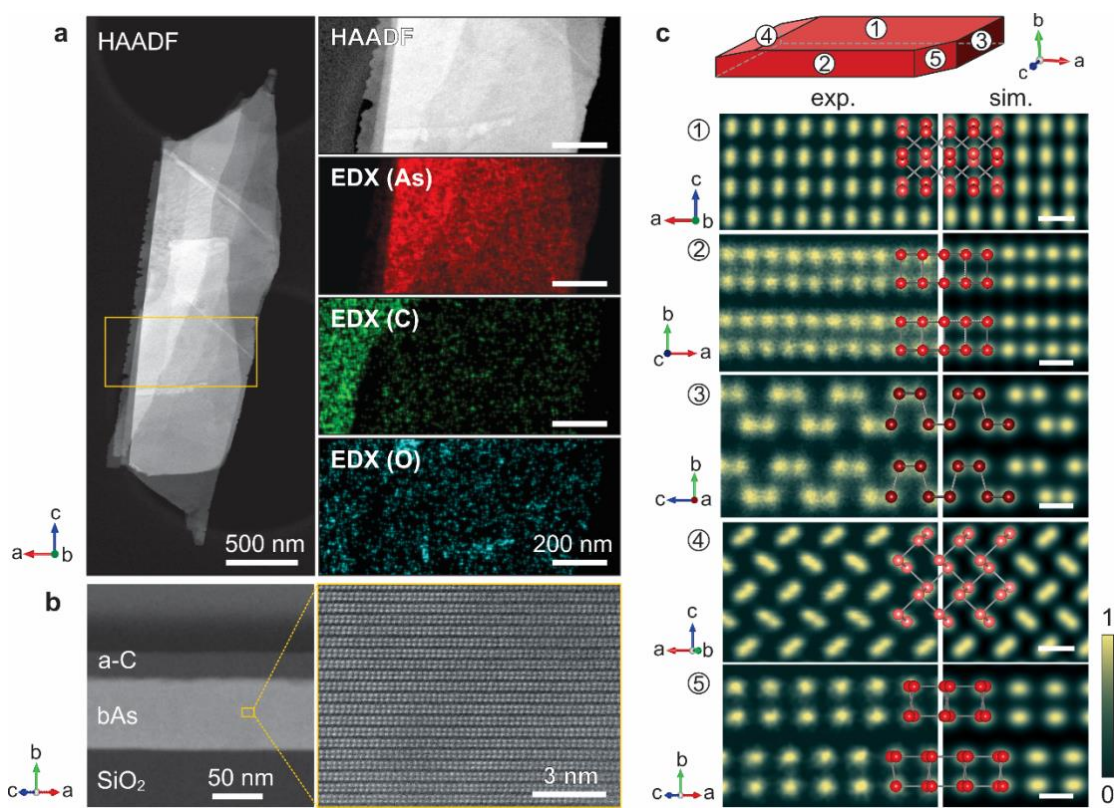

**Figure 1.** (a) Plan-view HAADF-STEM image of a BAs flake and EDX elemental maps obtained from a region of the flake (in yellow box) that is suspended over the hole in holey-carbon support TEM grid. $K_\alpha$ emission signals of As, C, and O were used for these elemental maps. (b) Cross-sectional HAADF-STEM images of a BAs flake deposited on Si/SiO$_2$ substrate in low- (left) and high- (right) magnifications. (c) Experimental and simulated atomic-resolution HAADF-STEM images of BAs from five different crystallographic orientations. A schematic on top illustrates the geometries of projection planes. Ball-and-stick atomic model of BAs for each orientation is shown on the HAADF-STEM images. Scale bars are 2 Å.

To examine the crystalline structure of BAs and determine the degree of anisotropy, atomic-resolution HAADF-STEM images of bulk BAs were obtained from different orientations, including three major crystallographic directions – armchair [001], zigzag [100], and plan-view [010] – and in two additional ones in minor directions ([110] and [101]). The results are summarized in Figure 1c (see SI, Figure S3 for details of image noise-reduction). The experimental data were compared with computed HAADF-STEM images simulated using the TEMSIM code,[20] which is based on the Multislice method.[21] Good agreement between the two, as shown in Figure 1c, confirms the assignments of the imaging directions (major: [010] (①), [001] (②) and [100] (③); minor: [110] (④) and [101] (⑤)). The experimental HAADF-STEM images were then utilized to evaluate the degree of anisotropy of BAs compare it to BP with a similar atomic structure. The lattice constant ratios (c/a, b'/a and b'/c) were obtained from the images of the three major axes (and tested using other two images for self-consistency) and compared to values for BP found in literatures (see Table 1). The ratios calculated using the theoretical lattice constants are also listed in Table 1 for comparison. The results, which are consistent with theoretical predictions, show that these BAs flakes are structurally highly anisotropic, but with slightly lower (~ 5%) in-plane anisotropy compared to BP.

**Table 1.** The lattice constant ratios for BAs and BP. Experimental values for BAs from this work is shown bold. a and c are in-plane lattice parameters in the zigzag and armchair directions and b' is half of the out-of-plane lattice parameter b.

|       | BAs (exp.)       | BAs (theor.)                         | BP (exp.)              | BP (theor.)              |
|-------|------------------|--------------------------------------|------------------------|--------------------------|
| c/a   | **1.24 ± 0.04**<br>1.166[5]<br>1.225[6]  | 1.264[22]<br>1.283[7]<br>1.259[8]    | 1.311[14]<br>1.355[2]  | 1.377[22]<br>1.378[23]   |
| b'/a  | **1.59 ± 0.08**<br>1.439[5]<br>1.523[6]  | 1.540[22]<br>1.543[8]                | 1.63[14]<br>1.668[2]   | 1.769[22]<br>1.678[23]   |
| b'/c  | **1.27 ± 0.06**  | 1.219[22]                            | 1.244[14]              | 1.285[22]                |

| | | | |
|---|---|---|---|
| 1.234$_5$ 1.243$_6$ | 1.226$_8$ | 1.231$_2$ | 1.223$_{23}$ |

*Thickness determination of a few-layer-thick BAs.*

The plan-view HAADF-STEM images can be used to precisely measure the thickness of atomically-thin 2D materials as the image contrast has a direct correlation with the number of atoms in a given atomic column. [24, 25] BAs has an AB stacking of layers where every other layer is half unit-cell shifted in the [100] direction from the layer before, as illustrated in Figure 2a.[14, 25] When viewed from the [010] direction, the lateral atomic position of the alternating layers, that are half unit-cell shifted in the [100] direction, can be seen. This configuration results in a distinct contrast in plan-view HAADF-STEM images for the odd and the even numbers of BAs layers.[14, 25] For the odd numbered BAs layers, neighboring atomic column-pairs (dumbbells) along the plan-view direction contain a different number of atoms, resulting in dissimilar intensities in a HAADF-STEM image. The intensity difference between adjacent atomic dumbbells is sensitive to the thickness of a BAs nanosheet, when the number of layers is relatively small. On the other hand, in case of even numbers of layers, the number of atoms in every dumbbell is identical, resulting in the almost same HAADF intensity for neighboring atomic columns (with very minor differences due to beam channeling).[26] The strong layer-dependence of the lattice contrast in HAADF-STEM images, for odd numbers of layers, can be utilized to directly measure the number of layers in the nanosheet. One such analysis, based on HAADF intensity ratios, is presented in Figure 2b-d, where the ratios from experimental images are directly compared with those from simulations to determine the number of layers in the edge region of a thin flake. It should be noted that the thickness estimation using the HAADF intensity ratio method is practical only in thin BAs nanosheets with less than about 15 layers (or < 9 nm), since the ratio saturates at the higher thicknesses (for more examples, see SI, Figure S4).

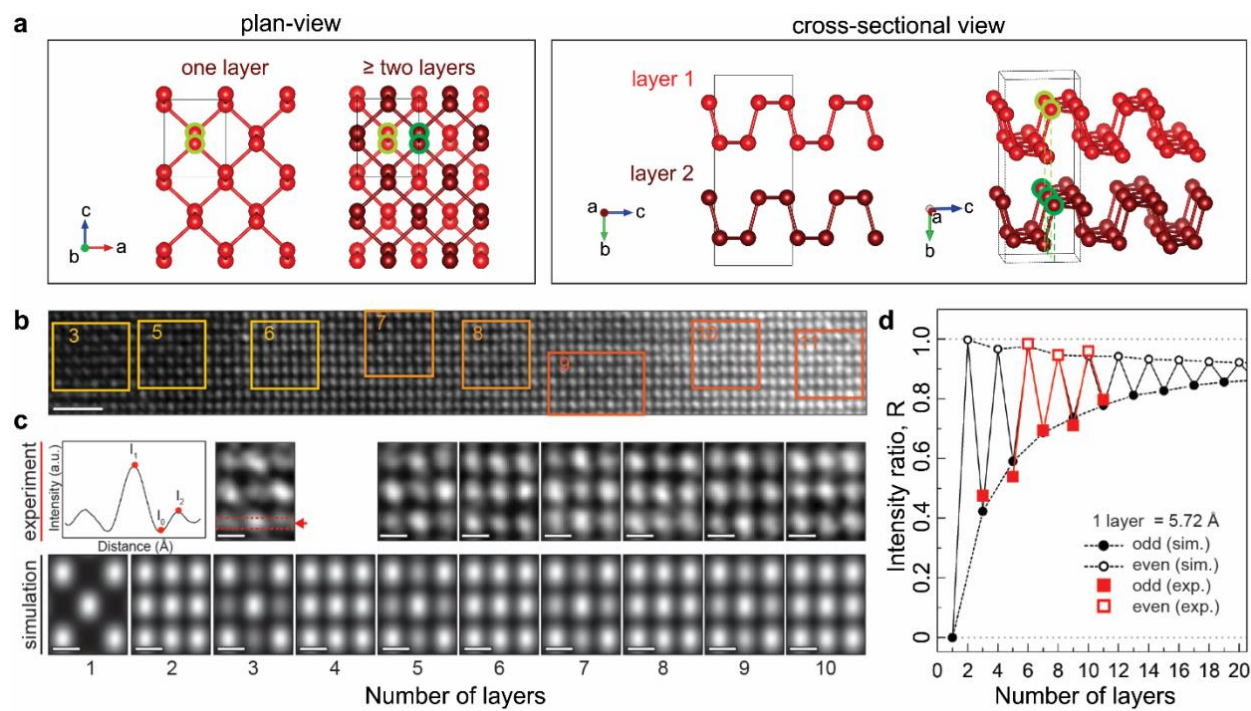

**Figure 2.** Layer-dependent lattice contrast of plan-view HAADF-STEM images. (a) Atomic model of AB-stacked structure of BAs. A unit cell is highlighted with the gray lines. Left: projected views of one layer-thick and two layer-thick BAs in the [010] (plan-view) direction are compared. Right: two layers in a unit cell are illustrated in the [100] (cross-sectional view) direction and slightly tilted view on the right illustrates the laterally shifted positions of atoms in every other layer. (b) HAADF-STEM image of an edge region of BAs flake displaying contrast variation due to the varying number of layers. The number of layers was estimated from selected regions. Scale bar is 1 nm. (c) Comparison of experimental and simulated HAADF-STEM images at different layer numbers. Sub-regions were selected from experimental HAADF-STEM image in (b) (marked with boxes). Scale bars are 2 Å. Line profiles were extracted from the region indicated with red arrows as shown on the top-left panel. The HAADF intensities between alternating sites ($I_1$ and $I_2$) are indicated. (d) HAADF intensity ratios, $R = (I_1-I_0)/(I_2-I_0)$, as a function of the number of layers. Intensity ratios from experimental images in different sub-regions are overlaid on reference ratios obtained from simulated images shown in (c).

*Dielectric response and electronic structure*

Low-loss and core-level EELS from BAs flakes were measured and analyzed to investigate the dielectric response and electronic structure of BAs. First, bulk EELS were acquired from relatively thick BAs samples (> 40 nm). Low-loss EELS, presented in Figure 3a, shows the bulk plasmon peak at $E_p$ = 18.6 eV and a series of features at the energies below $E_p$ due to interband transitions, Cherenkov radiation, and surface plasmon excitations.[27-30] While characteristic peaks originating from the lower and upper branches of surface plasmons in BAs are expected to be around 2 eV and 10–13 eV, as in BP,[14] they are not clearly identifiable here at this sample thickness because of strong "overlap" with other excitations. EELS measured using the higher energy resolution further reveals details of the fine structures in this lower energy-loss region (shown on the right). Three distinctly observed features are grouped as $a_1$, $a_2$, and $a_3$. A shoulder on the right-hand-side tail of the zero-loss peak (ZLP) is visible in the region $a_1$, which could be attributed to the band gap and surface plasmons. Band gap of bulk BAs is known to be ~ 0.3 eV,[5, 6] which is too small to be resolved using the available resolution of 0.13 eV and wide tails of the ZLP. Features $a_2$ and $a_3$, including two identifiable peaks at ~ 7.5 and ~ 8.8 eV, have the characteristics of interband electronic transitions and Cherenkov radiations; spectral shape does not vary in the thicker samples and scales with the bulk plasmon peak.

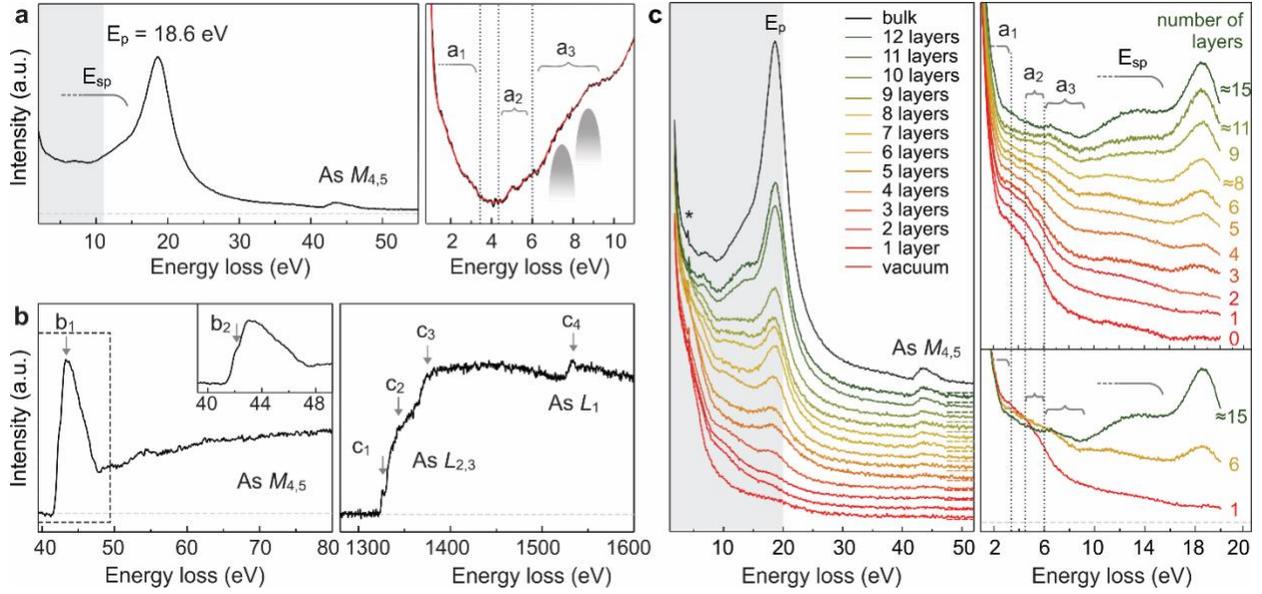

**Figure 3.** Experimental EELS data obtained from exfoliated BAs flakes. (a) Low-loss EELS with indicated bulk plasmon peak ($E_p$) and region of surface plasmon ($E_{sp}$) excitations. High energy resolution EELS from the range of 1–11 eV (shaded region) was acquired and shown on the right. The fine structures in the region are grouped and labeled as a1, a2, and a3. The positions of two peaks in a3 are indicated. (b) Measured core-loss EELS: As $M_{4,5}$ (left) and $L_1$ and $L_{2,3}$ (right) edges. Discernable peaks are labeled as b1 and b2, and c1 through c4. (c) Thickness-dependent low-loss EELS. (Left) Stacked EEL spectra that are vertically shifted for clarity and the artifact of measurement marked with asterisk. The higher energy resolution EELS obtained from the shaded energy range and shown on the right: stacked with vertical shifts (top) and a few of them shown without any vertical shifts (bottom).

**Table 2.** Peak positions in low-loss and core-level EELS shown in Figure 3.

|  |  | Energy (eV) |
| --- | --- | --- |
| Low-loss | a1 | 3.3 |
|  | a2 | 4.3 – 6 |
|  | a3 | 7.5, 8.8 |
| As $M_{4,5}$ | onset | 41.5 |
|  | b1 | 43.3 |
|  | b2 | 42.4 |
| As $L_1$, $L_{2,3}$ | onset | 1323 |
|  | c1 | 1326 |
|  | c2 | 1344 |
|  | c3 | 1375 |
|  | c4 | 1533 |

Core-loss EELS was also measured from BAs flakes, and the results are displayed in Figure 3b. The As $M_{4,5}$ edge with onset at 41.5 eV is from excitation of $3d_{3/2}$ and $3d_{5/2}$ core electrons to unoccupied *p* or *f* orbitals (selection rule: $\Delta l = \pm 1$) above the Fermi level.[30] The edge exhibits combination of both "saw-tooth" and "delayed maximum" shapes indicating that the core electrons excite to the unfilled bound electronic states (4*p* orbitals) near the conduction band minimum as well as continuum states at the higher energies.[30, 31] The As $L_{2,3}$ edge with onset at 1323 eV exhibits a typical "delayed maximum" edge from excitation of 2*p* core electrons to 5*s* and 4*d* orbitals. Distinct fine structures (labeled as $c_1$, $c_2$, $c_3$) are visible on top of strong tails of $L_{2,3}$ edges. The $L_1$ edge with a characteristic "saw-tooth" shape is also identifiable at 1533 eV (labeled as $c_4$). These features from measured core-level EELS are fingerprints of the electronic band structure for bulk BAs, as they are directly correlated with electronic density-of-states (DOS) above the Fermi level.[30, 32]

To investigate the dielectric response of BAs as a function of the layer number in thin BAs flakes, additional low-loss EEL spectra were acquired from the areas with different numbers of layers. In Figure 3c, a set of layer-dependent low-loss EELS, going from one to 15 layers with one-layer-steps, is presented. The exact layer number determination discussed above was used to evaluate the thickness of a particular area of the flake used for EELS (for details see SI Figures S5 and S6). It should be noted that due to long-range nature of the interactions for these low-energy electronic excitations,[30] some intermixing between these spectra is expected. As the layer number decreases from > 50 to below 20, the intensity of the bulk plasmon peak drops and the fine features at lower energy are modified. When the thickness of a BAs nanosheet is about 15-layer-thick, features due to surface plasmon excitations at around 12–14 eV starts to be visible. As the number of layers decreases to nine and lower, the features of surface plasmon losses are dampen and modified because of coupling of bottom and up surface plasmon oscillations and reshaping of the surface plasmon dispersion behaviour.[30] Similar in nature changes in surface plasmon dispersion in a few-layer BP have been reported.[33-35] Additional sets of spectra with higher energy resolution was also acquired in the energy-loss region below 19 eV (Figure 3c, on the right) to better highlight the changes discussed.

As the number of layers decreases to less than 6 layers, features in the range of $a_1$ and $a_2$ emerge. This can be seen in the bottom-right panel of Figure 3c, where the measured EEL spectra are displayed without vertical shifts allowing direct comparison of the intensities. The enhancement of the features in $a_1$ and $a_2$, below 6 eV, in such a thin BAs can originate from (1) strong coupling of surface plasmon modes enhancing excitations or (2) changes in the electronic band structure resulting in formation of high DOS near the conduction band minimum that are activated by electronic interband transitions. Recently, Zhong et al.[6] reported that as the number of layer is reduced, the band gap of BAs increases and carrier mobility near the conduction band minimum decreases, which implies that the onset of the low-loss region should be shifting to the higher energy and the DOS near the conduction band minimum should increase making the second argument plausible. These experimental results invite a detailed *ab initio* calculations of BAs dielectric function for different numbers of layers to provide a full understanding of the origins of these observed EELS fine structure changes at these low energies.

*Ambient stability*

Stability of BAs under different environments was also investigated. When an exfoliated BAs flake is kept at ambient conditions, a considerable structural destruction of the flake can be readily observed over relatively short time (in several days). An example of such ambient degradation of a BAs flake is presented in SI, Figure S7. Characterization of the degraded BAs flake, carried out by HAADF-STEM imaging and EDX elemental mapping, shows removal of As and no accumulation of oxidized As compounds (see SI, Figure S7), which is in contrast to the well-documented ambient degradation of BP, where formation of a:$P_xO_y$ dominates.[10-19] The removal of As during ambient degradation of the flakes suggests that destruction of BAs involves chemical reactions with atmospheric species that result in the formation of arsenic compounds other than arsenic oxides ($As_xO_y$), since arsenic oxides are solids at room temperature.[36] $H_2O$ in atmosphere, on the other hand, can participate in the degradation of BAs leading to formation of volatile products such as arsenic hydride (arsine, $AsH_3$ (g)) or arsenic acid ($AsO(OH)_3$ (aq)).[36, 37] Thus, the effect of $H_2O$ on the stability of BAs was studied by analyzing BAs flakes kept in humid and dry air. To rule out the photo-induced chemical reactions, the BAs flakes were kept in a controlled dark environment between STEM experiments, except during sample transfers into and from STEM during which the flakes were very briefly (~ 5 min) exposed to light and regular ambient conditions.

In Figure 4a, a time-series of HAADF-STEM images of the BAs flakes are presented showing distinct structural changes in the flakes under different conditions (for more examples, see SI, Figure S8). For direct comparison, images of each time-series were normalized to the same intensity scale using HAADF intensity of a supporting carbon film on a TEM grid as a reference. When BAs flakes were under humid environment (humidity of ~ 98%), destruction of most regions occurs just in 2 days indicating that moisture accelerates decomposition of the flake. Interestingly, a small region of BAs, that was not completely degraded on day 2, had a flat surface and several sharp boundaries along with eroded ones. On the other hand, flakes kept in dry air (humidity of ~ 2.5%, or practically without $H_2O$) did not show obvious change in HAADF-STEM images, indicating that $O_2$ alone does not decompose BAs.

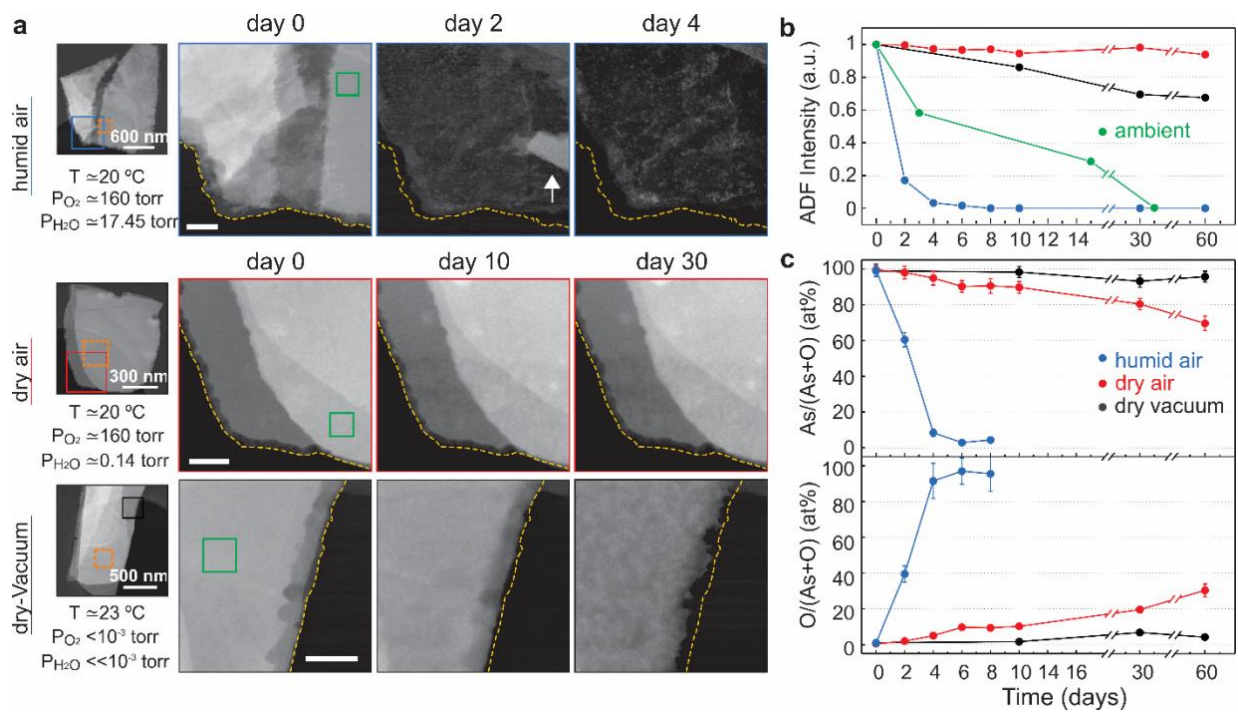

**Figure 4.** Degradation of exfoliated BAs flakes under different environments. (a) HAADF-STEM images of BAs flakes showing degradation over time. In the left most column, the images show the initial state of the flakes along with conditions they were kept. Orange box in the images show EDX acquisition area. The region in the solid box is magnified and presented at different times on the right. In the time-series images, original shape of flakes is outlined with dashed lines. The regions in the green box were used to measure the HAADF intensity. The initial thicknesses of the selected regions were between 15 and 20 nm. Scale bars are 50 nm. (b) Changes in HAADF intensity with time: humid condition–blue, dry condition–red, dry vacuum condition–black, ambient condition–green. (c) Relative elemental composition of BAs flakes as a function of time measured by EDX. Color code is the same as in (b).

The thickness change of the BAs flakes due to degradation was quantified using HAADF-STEM imaging. The average HAADF intensity of a region on the flake, indicated by green boxes in Figure 4a, was monitored with time and the results are presented in Figure 4b. As observed, the BAs flakes in humid air degrade over time, and it is with much faster rate than at ambient condition, while flakes kept in dry air exhibits negligible changes in the HAADF-STEM image intensity. The changes in chemical composition of BAs flakes during degradation, in particular, oxygen content were measured using EDX, as shown in Figure 4c. These EDX data were acquired from regions with the same thickness in each specimen highlighted with orange boxes in low magnification HAADF-STEM images shown in Figure 4a. Humid-conditioned BAs flake shows a systematic rapid increase of oxygen percentage due to vanishing of As. In the case of a dry air-conditioned BAs flake, the relative amount of oxygen also increases even though there is no reduction of thickness (or As content) according to the HAADF intensity. To further examine the role of oxygen on degradation, a set of experiments was conducted on BAs flake stored in dry vacuum condition (humidity of ~ 0%, P ≈ $10^{-3}$ torr). The results of these experiments are also presented in Figure 4 (for more examples see, SI Figure S8). Interestingly, dry vacuum-conditioned BAs flakes degraded in the same manner as the BAs at ambient conditions, but the amount of degradation is very minor (see Figures 4a and 4b). The degradation observed here is likely due to short exposures to ambient air when transferring, loading and unloading the sample into the microscope for analysis. In these experiments, the total sample exposure to ambient air was about the same as those in the other experiments. Considering the identical sample preparation procedure and similar ambient exposure of all samples, the observation of no obviously visible degradation in dry air-stored samples indicates formation of a protection layer on BAs in dry atmosphere. To rule out the effects of often occurring carbon contamination on dry-conditioned sample, the oxygen content of the BAs flakes stored in dry- and dry vacuum-conditions was assessed by analyzing elemental ratios of O/As and C/As (Figure 5a). In the case of dry air-conditioned BAs flake, oxygen content increases with time while carbon content stays unchanged. This suggests a formation of thin $As_xO_y$ layers on the sample surfaces under dry condition, which then acts as a protection layer during brief exposures of the sample to ambient condition before and after STEM analysis. It should be noted that dry air-conditioned flakes typically have slightly higher contaminations. Interestingly, a formation of a thin oxide layer on the surface of BP was also reported.[12, 15]

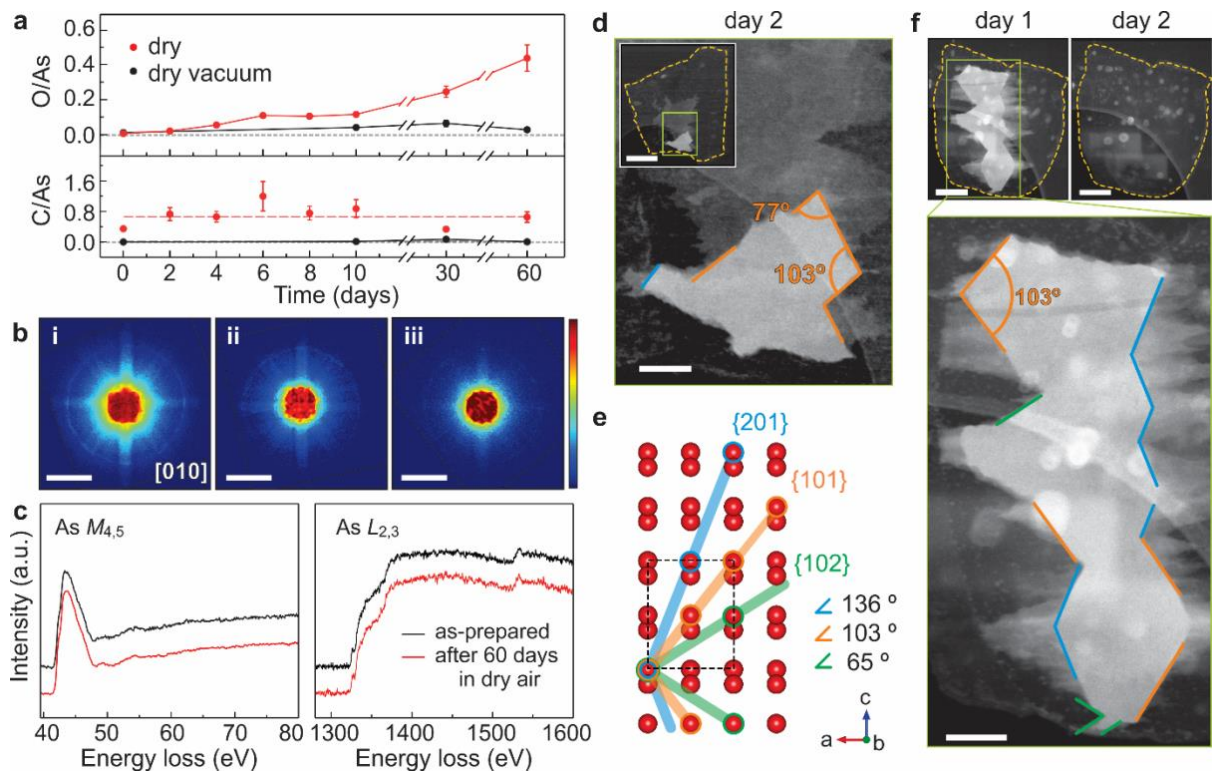

**Figure 5.** Analysis of degradation of exfoliated BAs flakes. (a) Elemental ratios (O/As and C/As) of BAs flakes in dry and dry vacuum conditions with degradation quantified by EDX. (b) CBED patterns obtained from as-prepared (i), dry vacuum-conditioned (ii), and dry-conditioned BAs (iii). The intensities in CBED patterns are displayed in log-scale. The scale bars are 2 Å$^{-1}$. (c) Comparison of core-loss EELS acquired from as-prepared and stored in dry air for 60 days flakes. (d) HAADF-STEM images of a small region of the BAs flake stored in humid environment for 2 days with a still mostly intact section. The edges and angles between them are indicated. Scale bar is 100 nm. The inset shows the entire flake with original edges outlined. The scale bar is 500 nm. (e) Atomic model of BAs illustrating {201} – blue, {101} – orange, and {102} – green crystalline planes from plan-view direction. A unit cell is shown by dashed lines. (f) HAADF-STEM images of a BAs flake stored in humid condition after being stored in dry air for 60 days. The original shape of the flake is outlined with dashed lines. A still mostly intact section of the BAs flake left on day 1 is shown in the bottom panel, where edges with identifiable lattice planes and angles are marked. The scale bar is 200 nm (top) and 100 nm (bottom).

To further understand the degradation of the structure, convergent beam electron diffraction (CBED) patterns of the flakes were also examined. Figure 5b shows CBED patterns acquired from dry- and dry vacuum-conditioned BAs flakes after being stored for 60 days which are compared with that from a fresh BAs flake (the initial flake thicknesses were about 30 nm). The CBED pattern measured from a fresh BAs flake shows clearly visible Kikuchi lines and the first order Laue zone ring, as would be expected from non-degraded sample, in a good agreement with the simulated CBED pattern (see SI, Figure S9). For dry vacuum-conditioned BAs flake the pattern is slightly weaker compared to the fresh sample, which would be consistent with the observed small thickness reduction. On the other hand, the pattern from dry-conditioned BAs flake is haze despite negligible thickness reduction. This would be consistent with presence of a thin amorphous $As_xO_y$ layer and some carbon contamination on the sample surfaces. Comparison of As $M_{4,5}$ and As $L_{2,3}$ core edges obtained from dry-conditioned and fresh BAs flakes (Figure 5c) shows no obvious spectral changes in their fine structures, again pointing out that thin $a:As_xO_y$ should be present only on the flake surfaces.

More insight into this water-enhanced degradation of the BAs flakes at ambient conditions can be gained from examination of the remaining sections of the flakes after a few-day degradation at ambient and humid conditions. HAADF-STEM images of remaining sections of BAs flake upon ambient exposure show non-directional degradation of the structure. The images of the small but still remaining fragments of humid-conditioned flakes show sharp edges and flat morphology, indicating directional (lateral) etching (see Figure 5d). For additional input, we also placed 60 days dry-conditioned BAs flake into humid conditions (as in previous experiments with fresh flakes) and studied its degradation. HAADF-STEM image of a remaining fragment is shown in Figure 5f, where again rapid etching with sharp edges can be seen similar to those observed for humid-conditioned flakes. Thickness reduction of the remaining fragments from humid-conditioned flakes with and without pre-dry-conditioning were evaluated using HAADF-STEM imaging and compared. Much slower thickness reduction from the top and bottom surfaces in the pre-dry-conditioned sample was noticed, which can be ascribed to the a:$As_xO_y$ layer on the surfaces protecting the flake surfaces from non-directional degradation.

The crystallographic planes of edges in the remaining fragment in the humid-conditioned flakes were identified. The observed angle of 77° (or complimentary 103°) between two edges correspond to unique angle between two crystalline planes of {101} (Figure 5e). Once these {101} planes are identified, using them as references, the planes corresponding to other edges can be also identified. The analysis of fragments indicates that etching preferentially takes place along the {101}, {201}, and {102} planes. Such directional etching is not completely surprising, as anisotropic etchings have been observed in other 2D materials including graphene,[40, 41] BN,[41] and $MoS_2$.[42, 43]

The result presented and discussed above demonstrate that both slower non-directional degradation and rapid directional etching of BAs nanosheets are facilitated by moisture in the air. The water-enhanced degradation at ambient condition suggests a few possible mechanisms. In one case, BAs readily reacts with oxygen molecules in the air, and the presence of water molecules facilitates subsequent reactions and promotes transformation of BAs into volatile arsenic compounds, such as $AsH_3$ or $AsO(OH)_3$. It is also possible that BAs directly reacts with $H_2O$ from edges and surface defect sites of the flakes, and the presence of oxygen in atmosphere further enhances the reaction promoting formation of volatile arsenic compounds. Detailed *ab initio* calculation-based study of the chemistry behind degradation of BAs is necessary to fully understand the degradation mechanisms and the exact role of each parameter.

In conclusion, 2D layered BAs was studied using STEM imaging combined with EDX and EELS spectroscopy. Atomic-resolution HAADF-STEM images acquired from five crystalline orientations confirmed the crystal structure and provide direct estimate of the structural anisotropy. It was also demonstrated that the lattice contrast in a plan-view HAADF-STEM image can by utilized to determine the number of layers in thin BAs nanosheets. The low-loss EELS, which is direct measure of the dielectric response of the material, were acquires from BAs nanosheets and shown to be very sensitive to the number of layers in the nanosheet. As the number of layers reduces below nine layers, the surface loss at 13–14 eV diminishes and a new feature at energy below 6 eV becomes dominant. The study of the stability of a BAs flake at different ambient conditions showed that, at atmospheric condition, BAs nanosheets degrade non-directionally and it is highly sensitive to moisture in air. At humid condition, BAs nanosheets additionally experience directional etching along {101}, {201}, and {102} crystalline planes. BAs flakes stored in dry air will form thin oxide layer on the surfaces and will be more resistant to non-directional degradation. The results presented here provide a guide for better utilizing the electronic and dielectric properties of BAs nanosheets and for improving their ambient stability. They will also play essential roles in evaluation of a-few-layer-thick BAs' full potential for incorporation into optical and electronic devices.

**Methods**

A. Sample preparation

Plan-view TEM samples were prepared by mechanically exfoliating bulk BAs using Scotch tape and then transferring them onto a poly-dimethylsiloxane (PDMS) stamp (Sylgard 184, Dow Corning Co.). The PDMS with the flakes was then stamped onto a 100-nm-SiO$_2$/Si substrate and then detached slowly leaving BAs flakes on the surface of the substrate. Next, polymethyl methacrylate (950 ka.u. PMMA C4, Microchem Co.) was spin-coated onto the substrate at 3000 rpm for 60 sec. followed by a soft bake at $120_o$C for 120 seconds. Then, $SiO_2$ was etched away in the etching solution (buffer oxide etchant 10:1 ($NH_4F$:HF)) leaving BAs/PMMA stacks floating over the etching solution. The BAs/PMMA films were washed using DI water and transferred to a TEM grid. Lastly, PMMA was washed off from the grid using acetone and left to dry in air for 1-2 minutes. For cross-sectional TEM samples, exfoliated BAs flakes on a $SiO_2$/Si substrate was sectioned using focused ion-beam (FIB) (FEI Helios Nanolab G4 dual-beam FIB) with 30 kV Ga-ions, which was further thinned with 2 kV Ga-ion beam to reduce the surface damaged layer. Amorphous carbon and Pt were sequentially deposited on the flake before the FIB-cutting to prevent damage from ion and electron beams of FIB. Specimen thicknesses were estimated using EELS log-ratio method with the plasmon mean-free path of $\lambda_P$ = 80.6 nm.[30]

B. STEM characterization

STEM experiments were performed using aberration-corrected FEI Titan G2 60-300 (S)TEM operated at 200 keV beam energy. The microscope is equipped with a CEOS DCOR probe corrector, super-X energy dispersive X-ray spectrometer, and a Gatan Enfinium ER EEL spectrometer. HAADF-STEM imaging and EDX elemental mapping were carried out using a beam current of ~30 pA and probe convergence angle of 17.2 mrad. ADF detector inner and outer angles for HAADF-STEM imaging were 55 and 200 mrad, correspondingly. STEM-EELS experiments were carried out using a monochromated STEM beam with a beam current of ~25 pA and probe convergence angle of 19 mrad. EELS detector acceptance angle was 29 mrad and the energy resolution was 0.13, 0.25, and 1.00 eV for energy dispersion of 0.01, 0.05, and 0.25 eV/channel. Energy resolution was determined from FWHM of the zero-loss peak. Very brief Ar plasma cleaning of specimens was carried out (for less than 7 sec.) before each experiment. No visible sample damage due to the electron beam exposure was observed under these STEM operational conditions.

C. Degradation experiment

Degradation experiments were conducted using three as-prepared plan-view samples. To prevent any photo-induced degradation, all samples were placed inside a light-tight box. The humid air-conditioned and dry air-conditioned samples were stored under at ambient pressure of 760 torr and temperature of 20 ± 0.2 ℃. Humidity level was controlled by locating deionized (DI) water and desiccants (Calcium sulfate purchased from Sigma-Aldrich, WA, US) placed alongside each sample in a closed-glass chamber, respectively (no direct contact with the specimens). Humidity and temperature of the closed-glass chamber were continuously measured (at minute intervals) with relative humidity maintained at 96–100 % and 0.6–5 % for humid and dry conditions, respectively. Since the vapor pressure of water at 20 ℃ is 17.5 torr, the partial pressure of $H_2O$ is estimated to be ~17.15 and ~0.44 torr for humid- and dry-conditioned samples, and partial pressure of oxygen is estimated to be ~160 torr (atmospheric condition) for both cases. Handling time for loading samples into the STEM chamber and for unloading was limited to be ~5 min to minimize exposure to light and ambient air. The dry vacuum-condition sample was stored in a vacuum box with desiccants. The vacuum level was measured to be in $1\times10^{-3}$–$2\times10^{-3}$ torr range and temperature was 23 ± 1 ℃.

D. HAADF-STEM image simulations

HAADF-STEM image simulation was carried out using the TEMSIM code [20] based on the Multislice approach [21] For the atomic structure of BAs model the lattice constants of 3.707 Å, 11.441 Å, and 4.686 Å in the [100], [010], and [001] directions were used with 8 As atoms per unit cell.[22] STEM probe parameters used were: $E_0$ = 200 keV, $C_{s3}$ = 0, $C_{s5}$ = 0, $\Delta f$ = 0, and $\alpha_{obj}$ = 17.2 mrad. The ADF detector inner and outer angles were 50 and 200 mrad, respectively, to match with experimental conditions. The slice thickness was set to be 1 Å for cross-sectional BAs models and 1.43 Å for plan-view models to preserve the atomic spacing. Frozen-phonon approximation[44] was used to include thermal diffuse scattering at T = 300 K. Root-mean square thermal displacement of 0.13 Å was used (see SI, Figure S9 for details). A source size of 0.8 Å was incorporated in these image simulations.[45]

E. Evaluation of column-to-column ADF intensity ratio in HAADF-STEM images

First, raw experimental HAADF-STEM images were low-pass filtered (0.6 Å) to remove the high frequency noise. Smaller sections of 1.39×1.39 nm (130×130 pixel) having uniform background were cut out and the intensities were individually renormalized for each section. Line scans were obtained across each row of the dumbbells averaging across 0.86 Å-wide strip to identify peak intensities $I_1$ and $I_2$ corresponding to that of neighboring dumbbells, and background intensity $I_0$. The simulated HAADF-STEM images are analyzed in an identical manner to obtain the ratio $R = (I_1-I_0/I_2-I_0)$ as a function of the number of layers.

## ASSOCIATED CONTENT

**Supporting information.**

- EDX spectrum of a BAs flake, image processing of high-resolution HAADF-STEM images, examples of thickness determination of BAs nanosheets using lattice contrast in atomic resolution HAADF-STEM image, number of layers determination for EELS acquisition area, BAs degradation at ambient condition, examples of BAs flake degradation, Pb planar defect in a BAs flake, local carbon contamination in a dry conditioned BAs flake, estimation of thermal displacement of BAs for the Multislice simulation, and Comparison of stability of pristine and pre dry-conditioned flakes. (BAs_supporting information.docx)

**Author Contributions**

H.Y., S.G., and K.A.M. conceived and designed the project and data analysis. H.Y. carried out all STEM sample preparation, STEM experiments, and diffraction pattern simulations. S.G. performed image simulations and thickness determination analysis. P.G. deposited flakes onto STEM grids and helped with preparation of cross-sectional samples. H.Y. and K.A.M. wrote the manuscript with contributions from all authors.

ACKNOWLEDGEMENT

This project was partially supported by UMN MRSEC program DMR-1420013 and SMART, one of seven centers of nCORE, a Semiconductor Research Corporation program, sponsored by NIST. Parts of this work was carried out in the College of Science and Engineering Characterization Facility, University of Minnesota (UMN), supported in part by the NSF through the UMN MRSEC program (No. DMR-1420013). P.G. and S.J.K. were supported by the NSF under Award No. ECCS-1708769. The authors also thank Sagar Udyavara and Prof. Matthew Neurock for insightful discussions.